# Self-enforcing Game Theory-based Resource Allocation for LoRaWAN Assisted Public Safety Communications

Vishal Sharma[1], Gaurav Choudhary[1], Ilsun You[1], Jae Deok Lim[2], Jeong Nyeo Kim[2]

[1] Department of Information Security Engineering, Soonchunhyang University, South Korea
[2] Electronics and Telecommunications Research Institute, South Korea
{vishal_sharma2012@hotmail.com, gauravchoudhary7777@gmail.com, ilsunu@gmail.com, jdscol92@etri.re.kr, jnkim@etri.re.kr}

## Abstract

Public safety networks avail to disseminate information during emergency situations through its dedicated servers. Public safety networks accommodate public safety communication (PSC) applications to track the location of its utilizers and enable to sustain transmissions even in the crucial scenarios. Despite that, if the traditional setups responsible for PSCs are unavailable, it becomes prodigiously arduous to handle any of the safety applications, which may cause havoc in the society. Dependence on a secondary network may assist to solve such an issue. But, the secondary networks should be facilely deployable and must not cause exorbitant overheads in terms of cost and operation. For this, LoRaWAN can be considered as an ideal solution as it provides low power and long-range communication. However, an excessive utilization of the secondary network may result in high depletion of its own resources and can lead to a complete shutdown of services, which is a quandary at hand. As a solution, this paper proposes a novel network model via a combination of LoRaWAN and traditional public safety networks, and uses a self-enforcing agreement based game theory for allocating resources efficiently amongst the available servers. The proposed approach adopts memory and energy constraints as agreements, which are satisfied through Nash equilibrium. The numerical results show that the proposed approach is capable of efficiently allocating the resources with sufficiently high gains for resource conservation, network sustainability, resource restorations and probability to continue at the present conditions even in the complete absence of traditional Access Points (APs) compared with a baseline scenario with no failure of nodes.

## 1 Introduction

Public Safety Communications (PSCs) are the important pillars of society that play a paramount role at the time of emergency. In case of critical situations like natural disasters, fires, terrorist attacks and a major contingency, PSC carries the entire communication load on their shoulders. In such kind of incidents, the emergency management services like disaster recovery teams, government bodies and emergency medical accommodations play a pivotal role for controlling the critical situations and support public safety operations to preserve lives and infrastructures. It has been increasingly apperceived that efficacious communications are the key to successful management of emergency and disaster situations [1-2]. Therefore, the stability of PSC in a very arduous environment should be reinforced where the availabilities of alternative infrastructure and services are missing. This task requires a set of functional capabilities, which include low-complex data acquisition, resource allocation and management, and secure and reliable communications.

Public safety networks are the providers of a shared medium among many organizations in a traditional or ad hoc format. In case of emergencies, the active devices compete for available resources. Consequently, there are many users and it is very consequential to allocate shared resources in a very fair and constrain manner to consummate the requisite of each utilizer [3-5]. Furthermore, the available resources are constrained and their efficient utilization is of primary concern. There are several resource allocation schemes existing in the literature, but the majority of them fail to meet Quality of Services (QoS) requisites along with the availability and the reliability for communication for public safety users. These available resources can further be utilized with a fair scheduling and optimal allocation strategies [6-7].

PSC relies on existing wireless network setups, which are the backbone in case of mobility and astronomically immense functionality of resources. Public safety networks can further be utilized to operate in coordination with the Internet of Things (IoT) as well as with certain cyber-physical entities for reliable and continuous transmissions. However, there are major challenges associated with such combination-based PSC networks, which include policy control for the involved devices, security of IoT and PSC servers, resource restoration, consistent data acquisition, logging, energy conservation, localization, community or node structure maintenance, privacy-preservation, etc [8-12]. All of these are research quandaries and need effective strategy by means of novel network architectures and protocols.

Recently, many organizations and academia have fixated on the development of broadband public safety networks predicated on LTE and 5G [13-15]. But in case of emergencies like natural disasters or maleficent attacks, the mobile telecommunications may breakdown that may halt the communication services for controlling the threatening effects and consequences due to rumors transmitted over the active network. Such a scenario requires analysis and deployment of an efficacious strategy that can be facilely managed and controlled without affecting the availability and operational-activity of the already operating devices [16-17].

To address this issue, LoRaWAN, with the property of secure bidirectional communication, mobility and localization, is considered as a significant solution. LoRaWAN has become one of the most significant technologies due to its property of long-range communication with energy efficient computations [18-19]. Its protocol and network architecture directly influence the battery life of a node, network capacity, QoS, security, and the variety of applications served by the network. It requires 5 to10 times less number of Base Stations (BS) than 3G/4G and can be deployed for networks with dynamic nodes as BSs. An overview of LoRaWAN is presented in Figure 1 [20-21]. This paper uses a combination of LoRaWAN and traditional cellular setup while solving issues related to resource allocation between their servers for PSCs.

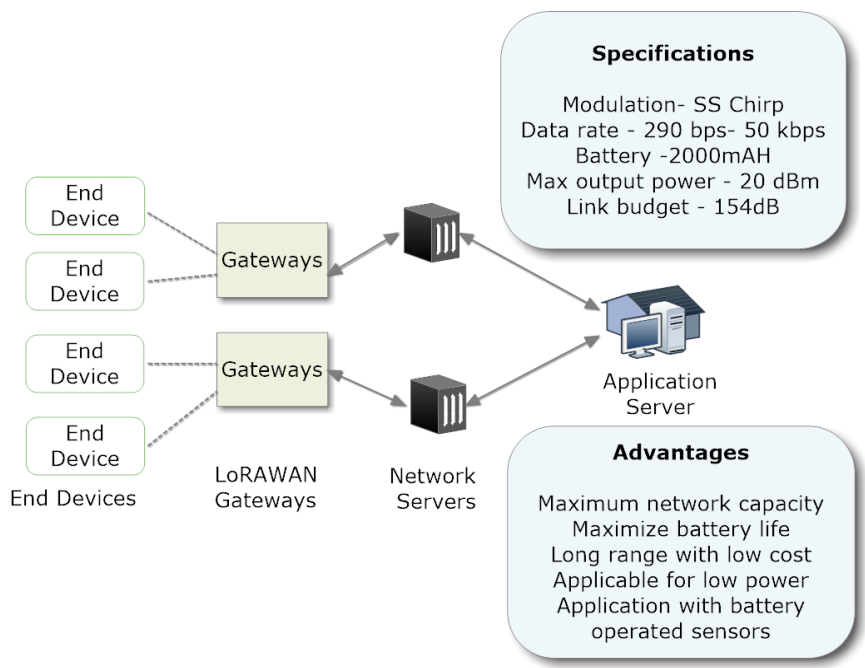

**Figure 1.** A general overview of the LoRaWAN architecture with its specifications and advantages

## 1.1 Problem Statement

Traditional networks provide separate infrastructural support for handling public safety applications and use designated entities for regulating the PSC traffic. However, if these networks are unavailable due to any unintentional circumstances, the peril of failing for PSC increases. This may engender havoc like situation leading to immensely colossal losses in terms of services and users. Such a scenario can be fortified by shifting traffic to a secondary network and letting PSC operate without fail. However, a constraint of resources and over-burden may again fail the entire setup. Thus, regulation of traffic on the basis of available resources is of utmost importance. Similarly to such requisite, this paper considers LoRaWAN as the secondary network that avails to keep traffic emanating from PSC sources live and regulates them across the network. However, this shifting of applications leads to overconsumption of own resources for LoRaWAN, which is a problem at hand. It is desired that the infrastructure of LoRaWAN should cope with an excessive burden of applications from traditional setup of public safety networks, but should not compromise its own traffic, nor does should defy any requests from its own applications.

## 1.2 Our Contribution

This paper addresses the issue of resource allocation, which arises once the traditional networks that are designated for PSCs are unable to process the incoming traffic. The major challenge is to use the alternatively available network for supporting excessive applications without affecting its own performance. One such combination of LoRaWAN and traditional networks is used in this paper for supporting PSC in case of a scenario similar to a disaster. The proposed approach forms self-enforcing agreement [22] via game theory modeling and uses Nash equilibrium [23] to allocate resources efficiently among the available network servers. The proposed approach emphasizes the memory and the energy as key metrics and provides sufficiently high sustainable and survivable network architecture with a high resource restoration mechanism by using LoRaWAN servers in form of a virtual cloud. The evaluations of the proposed approach are done numerically and the performance is evaluated in comparison to a baseline scenario with no failures.

The remainder of the paper is structured as follows: Section 2 presents a study on existing works for resource allocation and PSCs. Section 3 presents the initial setup and the proposed network model. Section 4 provides details of using self-enforcing agreement along with Nash equilibrium for the defined system model. Performance evaluations are presented in Section 5. Section 6 compares the proposed approach with the state-of-the-art solutions. Finally, Section 7 concludes the paper.

## 2 Related Works

PSC has a higher priority in terms of availability, coverage area and reliability. Incident response team performance relies on PSC because of its efficiency which is directly proportional to the performance of incident response organizations. Therefore, many governments are heavily investing for its nation-wide development. There are several schemes for PSC, but many of them lack behind in terms of availability of infrastructure during a disaster. An overview of related works is presented in Figure 2.

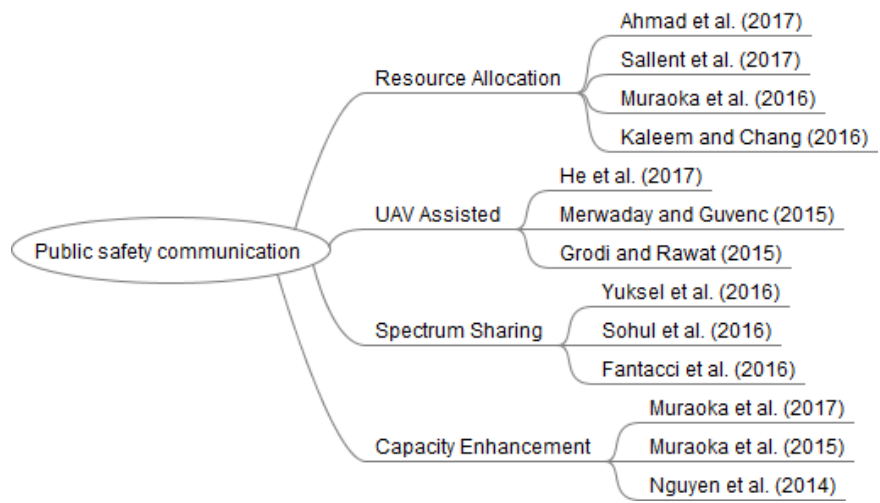

**Figure 2.** An overview of the related works for public safety communications

During the last decade, researchers have continuously focused on PSC to improve security, dependability, fault tolerance, cost-effectiveness, interoperability, spectral efficiency, and advanced capabilities of existing networks during a disaster. Resource allocation has been a dominating concern for PSC since the majority of the applications operate as a service hungry entity leading to congestions, degradation in QoS, or even failure of the entire network. There have been a lot of solutions for resource allocation in case of disaster communications. In the recent time, Ahmad et al. [24] provided co-channel interference analysis for the coexisting public safety and railway networks by using cooperative interference coordination schemes. This scheme supports the user priority-based cooperative resource allocation in Long Term Evolution (LTE). LTE for PSC can deliver mission-critical broadband data with minimum latency. But both the two networks that both comply with the LTE standards can have different ways of handling roaming.

Kaleem and Chang [25] developed a new public-safety priority-aware user association scheme for interference reduction and load balancing in a PSC-LTE system. The proposed scheme improves the user-association problem by minimizing the Call Blocking Probability (CBP) w.r.t. the network load conditions and PSC users. Chen et al. [26] proposed hybrid network architecture on the resource allocation in public safety broadband network with a rapid deployment of AP in the LTE network. Furthermore, many schemes incorporate LTE with resource allocations to improve the PSCs. Muraoka et al. [27] proposed methods for allocating radio resources using scheduling for D2D communications. Through their approach, scheduling can be used for efficiently handling local traffic both in normal and large cells. But load balancing and resource sharing are yet to be resolved for these networks as stated in Sallent et al. [28].

The concept of interleaved resource mapping for autonomous D2D discovery in public safety LTE, in terms of capacity enhancement, is coined by Ohtsuji et al. [29]. Furthermore, the feasibility of capacity enhancement for public safety is covered in Muraoka et al. [30]. For effectively applying LTE in PMR context, Nguyen et al. [31] provided a new architecture for multi-users multiplexing radio voice transmissions over LTE. The spectrum is shared among user for enhancing the availability of public safety networks. Yuksel et al. [32] suggested multi-hop and multi-technology pervasive spectrum sharing (PSS) concept. Sohul et al. [33] discussed the value of spectrum sharing for supporting the long-term vision in PSC. Fantacci et al. [34] focused on critical issues that impact PSC evolution towards new technologies via spectrum sharing. But it is difficult to pinpoint accessible spectrum bands without hurting their licenses operations, which is a challenging task in itself. As a solution for dynamic resource allocation, recently, many researchers have valued the importance of using on-demand and dynamic nodes, such as drones, as a crucial entity in PSCs. Most of them have suggested on using drones as a communication gateway that can support incoming and outgoing connections. Drones can be configured as a regular network entity and can be used similarly to already operating network units [35-39]. Grodi and Rawat [40] designed a UAV-assisted wireless network to provide connectivity during disasters. In their work cloud-based distributed database centre monitors the overall network and provides feedback to the emergency response centers when needed. However, the biggest challenge is to maintain a tradeoff between coverage area and propagation delay when deploying UAV networks for PSCs.

# 3 Network Model

This section describes the initial and the LoRaWAN-assisted PSCs along with the problem formulation.

## 3.1 Initial Setup

The network comprises a macro cell-BS serving as the core network server of the LoRaWAN and is connected to the core of the network, which is serving as the application server from where all the application interfaces are available to the end user. The application server is vendor-specific and is operated as per the configurations and decisions of the service providers. The number of core network servers for the entire area is represented by the set $N$, each of which is further connected to a set $S$ of APs. Each AP is responsible for managing core network services and operates over a high-speed link. Further, the core network server or BS is divided into multiple sub network servers, which are denoted by a set $B$. Each network server can be connected to multiple LoRaWAN gateways that help to connect the users to the application server via sub-network servers and core network server. The LoRaWAN gateways are represented by a set of $L$ and the end users are marked by a set $E$. At the moment, the network supports application specific data where each LoRaWAN gateway operates for applications related to data operation such as agriculture, industrial sensing, medical facilities notification and surveillance. Note that the APs in the network support regular traffic and helps to disseminate cellular traffic across the network. Further, the core network server only takes into account the management of sub-network servers of LoRaWAN setup. It does not account for LoRaWAN operation in the initial or regular connectivity and act as a mere router for LoRaWAN traffic, the public safety network is operated via regular transmission over cellular setup through multiple APs. In the initial phase, the network operates independently and has a limited dependency on LoRaWAN setup. However, in the scenario where APs cannot be accessed for PSC, the network should be initiated for comprehensive operations through the available resources. Such a scenario is described in the next sub-section on the basis of which the proposed solution is developed.

## 3.2 LoRaWAN-assisted Network Setup

LoRaWAN assistance is required when the regular network is unable to support the normal as well as critical transmissions, as that of PSCs during a disaster-like scenario. This part considers such a case when limited support is available from the regular devices and the network has to depend on LoRaWAN setup for PSCs making sure that the actual traffic of LoRaWAN is unaffected. Such a problem is similar to the resource allocation and job migration between the regular and the LoRaWAN devices.

Let assume that out of $|S|$ APs, $X$ is the number of APs are unavailable so the set of available APs can be given as $S'$, such that $S' \subseteq S$ and $|S'|=|S|-X$. Assuming that there is a full support from the LoRaWAN setup and there is no effect on the communication of its devices even during a disaster as shown in Figure 3.

Now, the communication support for PSC via LoRaWAN infrastructure is subject to the solution of resource allocation problem which helps to maintain regular operation intact and unaffected. The network model operates by identifying the resources and performance metrics on the basis which the entire resource allocator is formulated. These models include:

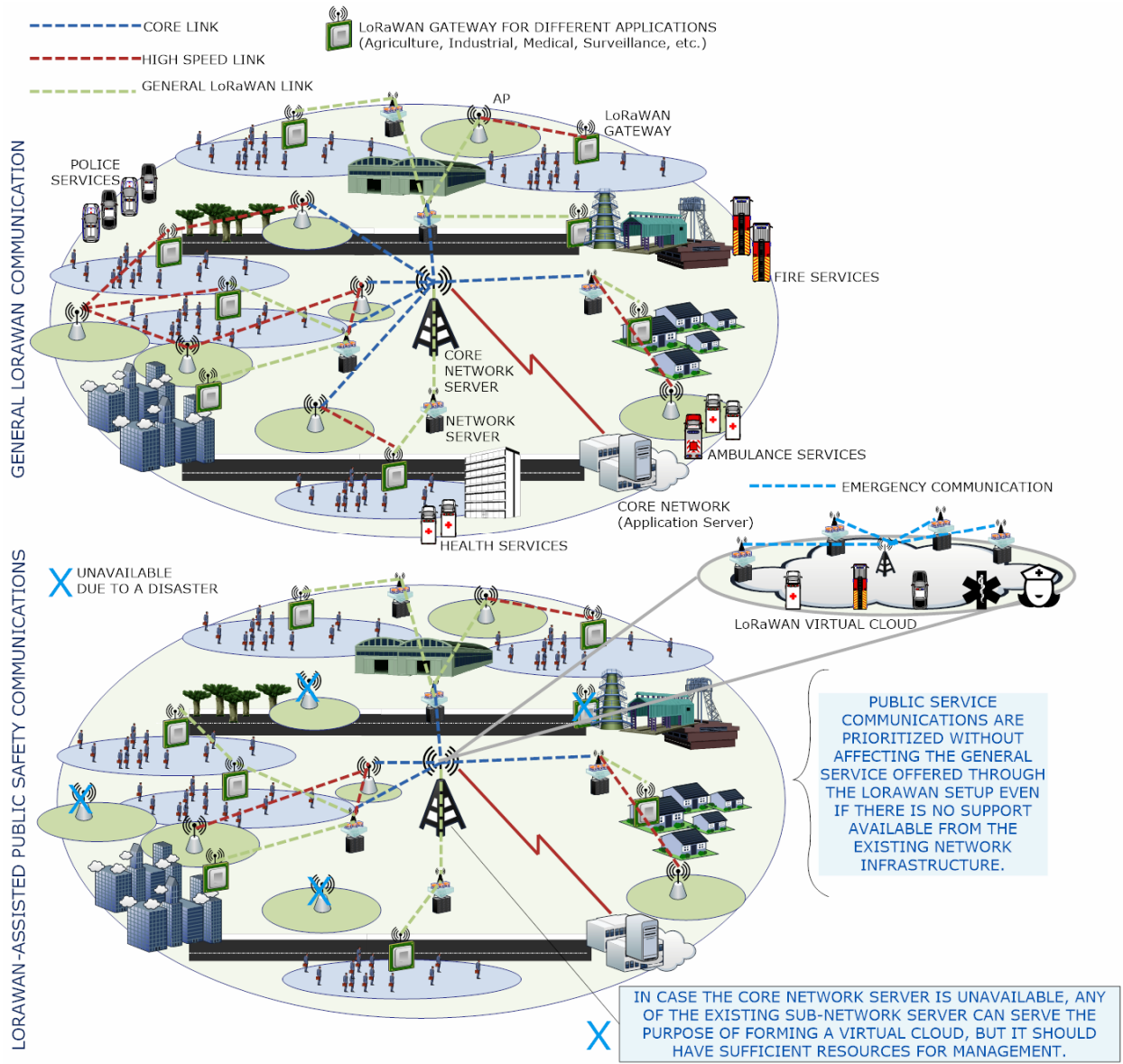

**Figure 3.** An exemplary illustration of general network architecture and the proposed LoRaWAN-assisted PSCs during a disaster like scenario

**Web traffic rate.** It is the incoming rate of the requests from the LoRaWAN and the PSC end users that are to be shared over the network server towards the core.

**Application dominance ratio.** It is the priority ratio of the requests coming at the same instance. The priority ratio is calculated as the number of slots requested by LoRaWAN traffic to the total slots available. Similarly, application dominance ratio can be calculated for PSC application and the one that satisfies the conditions imposed by the proposed solution wins and has highest dominance ratio.[1]

**Return time and turnaround time.** Return time refers to the time consumed in getting final confirmation for an application once it is submitted; whereas the turnaround time is the time to obtain the first acknowledgment for an application. Any application processing is subject to dominance and the ones with the same dominance depend on the scheduling criteria which uses return time and the average turnaround time for decisions on allocation.

Now, for resource allocation, self-enforcing agreement [22] approach is used for defining the problem and the solution is enforced while solving the defined problem. The only issue is to fit in the extra traffic requests from the PSC network. Thus, the system model is formulated for PSC assuming that the readings for required metrics that are defined for PSC are already available for LoRaWAN setup.

Let λ be the arrival rate of traffic from the PSC end device with *K* number of applications. Each application request coming with this arrival rate is subject to memory and energy requirements denoted by α and β, respectively. For this, the memory requirements are calculated as:

---

[1] Although the assumption is strong, but it helps to understand the full capacity with which the LoRaWAN can support PSCs.

$$\alpha_T^{(M)} \le \sum_{i=1}^{K} (\alpha)_i \text{, such that } \alpha_T^{(M)} \le \alpha_A^{(L)}, \quad \textbf{(1)}$$

where $\alpha_A^{(L)}$ is the available memory for the LoRaWAN setup. Further, at any instance, the allocation condition follows,

$$\alpha_i \le \alpha_N^{(L)}, \quad \textbf{(2)}$$

where $\alpha_N^{(L)}$ is the available memory for the initial network server to which the PSC is intended for transmissions over the network. Let $T$ be the total time for which the entire session of PSC should last and $t_1^{(\alpha)}, t_2^{(\alpha)}, t_3^{(\alpha)} \dots t_B^{(\alpha)}$ be the time slots for which the memory will last for $|B|$ number of network servers as per the present rate $\lambda$. Now the probability that the memory will be available for at least $t$ time is given as:

$$\frac{1}{t} \sum_{i=0}^{t' < t} (P)_i, \quad \textbf{(3)}$$

where $P_0, P_1, P_2, \dots, P_t$ are calculated as the ratio of the required memory to the available memory, such that required memory is always less than equal to the available memory. If at any stage, this condition is violated, the system cannot operate any further and it is the condition of congestion or deadlock. Let $\Delta\alpha$ be the excessive memory available for each of the incoming traffic, then the expected gain at each network server will be given as [22]:

$$G_{ain}^{(P)} = \sum_{i=0}^{T} (\Delta\alpha)_i, \quad \textbf{(4)}$$

and the total gain of the network will be given as:

$$G_{ain}^{(N)} = \sum_{i=1}^{|B|} \left( \sum_{j=1}^{K} \left( \sum_{k=1}^{T} (\Delta\alpha)_k \right)_j \right)_i, \quad \textbf{(5)}$$

$$G_{ain}^{(N)} = \sum_{i=1}^{|B|} \left( \sum_{j=1}^{K} \left( G_{ain}^{(P)} \right)_j \right)_i. \quad \textbf{(6)}$$

Now, from the theory of self–enforcing agreement [22], the expected value of gain for all duration is given as:

$$E\left(G_{ain}\right) = \sum_{i=1}^{K} \left( P_{T+1} G_{ain}^{(N)} \right)_i, \quad \textbf{(7)}$$

$$= \sum_{i=1}^{K} (Q_T \Delta\alpha)_i. \quad \textbf{(8)}$$

Here, $Q_T$ denotes the probability of memory condition on the basis of which an approach or an application can be operated for a time period greater than $T$. Let $\Delta\alpha_x$ be the gain observed by each network server once the PSC applications are unavailable, because of which the total gain becomes $\Delta\alpha_x\alpha_x$. Now the reverse phenomenon on self-enforcing agreement will give the condition to continue with the current allocation policies [22]. According to which,

$$E\left(G_{ain}^{(N)} + \Delta\alpha_x\right) = E\left(G_{ain}^{(N)}\right) + Q_T\Delta\alpha_x - \sum_{T+1}^{T_{max}} (Q_T\Delta\alpha). \textbf{(9)}$$

By using the definition of self-enforcing agreement, the conditions for decisions are marked as:

$$\left[ E\left(G_{ain}^{(N)} + \Delta\alpha_x\right) - E\left(G_{ain}^{(N)}\right) \right]^R =$$
$$\begin{cases} Q_T\Delta\alpha_x - \sum_{T+1}^{T_{max}} (Q_T\Delta\alpha) < 0 \{\text{Reset}\} \\ Q_T\Delta\alpha_x - \sum_{T+1}^{T_{max}} (Q_T\Delta\alpha) = 0 \{\text{Continue}\} \\ Q_T\Delta\alpha_x - \sum_{T+1}^{T_{max}} (Q_T\Delta\alpha) > 0 \{\text{Reallocate excess memory for PSC}\}. \end{cases} \quad \textbf{(10)}$$

The network model is further modeled for energy condition, which is another major metric for deciding the policies for resource allocation. For the given network, let $\varepsilon_X^{(L)}$ be the energy consumed at the network server for LoRaWAN such that:

$$E_X^{(L)} = \sum_{i=1}^{K} (P_C \times t)_i, \quad \textbf{(11)}$$

which can be scaled to entire network as:

$$\sum_{i=1}^{|B|} (E_X^{(L)})_i. \quad \textbf{(12)}$$

This consumption of energy should be lower than the degradation rate [42] for available energy. This can be expressed as:

$$E_X^{(T)} - E_X^{(L)} \le \left( E_0 e^{-KT} \right), \quad \textbf{(13)}$$

where $E_0$ is the initial energy of the network, and initially $\varepsilon_X^{(L)} = E_0$. Now, the energy requirement for PSC should also satisfy the condition given above. By using policies of the self-enforcing agreement [22], let $\Delta E$ be the energy gain available at each network server and $P_E$ be the probability that the network can survive for $T$ duration at present decay rate [22] [42], such that:

$$\sum_0^T P_E = 1, \quad \textbf{(14)}$$

which implies

$$Q_E = \sum_{T+1}^{T_{max}} P_{E(T)}, \quad \textbf{(15)}$$

where $Q_E$ is the probability such that $\Delta E$ will be sufficiently large to continue PSC transmissions ever after $T$ duration and $\varepsilon_P^{(R)}$ is the required energy for one instance, such that for each application at a particular network server for higher $P_E$,

$$\sum_{i=1}^{T}\left(E_P^{(R)}\right) \le \Delta E,\qquad (16)$$

and for higher $Q_E$

$$\sum_{i=1}^{T}\left(E_P^{(R)}\right) << \Delta E.\qquad (17)$$

Now, the expected value for conservation of energy is given as:

$$E\left(E_{gain}\right)=\left(\sum_{i=1}^{K}(P_E)\right)\Delta E$$
$$= Q_E \Delta E\qquad (18)$$

From the definition of self-enforcing agreement, and applying reverse-strategy, the conditions for continuity w.r.t. energy are given as [22]:

$$E\left(\Delta E + \Delta E_X\right)=E(E_{gain})+Q_E\Delta E_X - Q_E\Delta E,\ (19)$$

and

$$\left[E\left(\Delta E + \Delta E_X\right)-E\left(E_{gain}\right)\right]^R =$$
$$\begin{cases} Q_E\Delta E_X - Q_E\Delta E < 0\{\text{Reset}\} \\ Q_E\Delta E_X - Q_E\Delta E = 0\{\text{Continue}\} \\ Q_E\Delta E_X - Q_E\Delta E > 0\{\text{Allow more applications for load balancing}\}\end{cases}\qquad (20)$$

Both the memory and energy equations help to manage the network during application requests from PSC setup. However, both these metrics depend on other parameters for a mutual decision on the resource allocation. These metrics are resource restoration, network useful time and network survivability. In the given network, resource restoration can be calculated as:

$$R_S^{(N)}=\left(\frac{K'}{K}\times\left\|\frac{L_A}{L_T}\right\|\right)\left(\frac{n_1\|E_R\|+n_2\|\alpha_R\|}{n_1+n_2}\right),\qquad (21)$$

where $n_1+n_2 > 0$, $n_1,n_2\le 1$, $n_1$ and $n_2$ are the balancing constants for energy and memory, respectively. $K'$ is the number of PSC applications closing at a given instance. $L_A$ is the number of links active on a given network server and $L_T$ is the total number of links. $\|E_R\|$ and $\|\alpha_R\|$ are the normalized value for energy and memory restorations, respectively, which are calculated as:

$$E_R=\frac{\sum_{i=1}^{K'}(P_C\times t)_i+\Delta E_R^{(M)}}{E_0}-E_0 e^{-KT}.\qquad (22)$$

The normalization is performed for all network server such that:

$$\|E_R\|_{|B|}=\left\|\frac{\sum_{i=1}^{K'}(P_C\times t)_i+\Delta E_R^{(M)}}{E_0}-E_0 e^{-KT}\right\|_{|B|}.\ (23)$$

The above formulation can also be expressed trivially, such that:

$$\|E_R\|_{|B|}=\left\|\frac{\sum_{i=1}^{K'}(P_C\times t)_i-\sum_{j=1}^{K}(P_C\times t)_j+\Delta E_R^{(M)}}{E_0}\right\|_{|B|}\qquad (24)$$

Here, $\Delta E_R^{(M)}$ is the average remaining energy of the network server at all time. Similarly, the memory restoration rate can be expressed as:

$$\alpha_R=\frac{\sum_{i=1}^{K'}(\alpha)_i-\Delta\alpha_R^{(M)}}{\alpha_A^{(L)}}-\alpha_0 e^{-KT},$$

$$\text{Or}\qquad (25)$$

$$\|\alpha_R\|_{|B|}=\left\|\frac{\sum_{i=1}^{K'}(\alpha)_i-\Delta\alpha_R^{(M)}}{\alpha_A^{(L)}}-\alpha_0 e^{-KT}\right\|_{|B|}$$

For simplicity $\alpha_0=\alpha_A^{(L)}$ Also, like energy formulation, memory restoration rate can be written trivially as,

$$\|\alpha_R\|_{|B|}=\left\|\frac{\sum_{i=1}^{K'}(\alpha)_i+\Delta\alpha_R^{(M)}-\sum_{j=1}^{K}(\alpha)_j}{\alpha_A^{(L)}}\right\|_{|B|},\qquad (26)$$

where $\Delta\alpha_R^{(M)}$ is the average remaining memory of a network server at all the time. In the given network, useful time of a particular network server should be cooperated for handling the excess of PSC applications. This will help to balance the load and also help to sustain in case the energy and the memory conditions are violated. The network useful time is expressed as the time up to which the useful slots available can be extended, which can be calculated as the mean lifetime, such that:

$$\Gamma=\left(\frac{1}{K-K'}\right),\qquad (27)$$

$$\Gamma'=S_0\Gamma,\qquad (28)$$

where $S_0$ is the number of slots available for a particular network server. Now, the network survivability can be written as:

$$S_N^{(R)} = \begin{cases} \Delta E \geq \sum_{i=1}^{K''} (P_C \times \Gamma)\{Available\} \\ \Delta E > \sum_{i=1}^{K''} (P_C \times \Gamma)\{Unavailable\} \end{cases}, \quad (29)$$

where $K''$ is the remaining applications. Now, for the entire session, the average sustainability can be given as:

$$S_{avg_1}^{(R)} = \frac{1}{T_{max}} \int_0^{T_{max}} E(G_{ain})(t)dt, \quad (30)$$

$$S_{avg_2}^{(R)} = \frac{1}{T_{max}} \int_0^{T_{max}} E(E_{gain})(t)dt, \quad (31)$$

$$S_{avg}^{(R)} = \frac{1}{2}\left(S_{avg_1}^{(R)} + S_{avg_2}^{(R)}\right). \quad (32)$$

From the above equations, the problem for the resource allocation can be expressed as:

$$\begin{aligned} &\max\left(R_S^{(N)}\right), \forall B, \forall L, \forall E, \\ &\max\left(\Gamma'\right), \forall B, \forall E, \\ &\max\left(S_N^{(R)}\right), \forall B, \forall L, \forall E. \end{aligned} \quad (33)$$

s.t.

$$\begin{aligned} &P_E = \max, Q_T = \max, \\ &\Delta E_X = \max, \Delta\alpha_X = \max, \\ &E_R \geq 0, \alpha_R \geq 0, \\ &\max\left(\left[E(\Delta E + \Delta E_X) - E(E_{gain})\right]^R\right), \\ &\max\left(\left[E\left(G_{ain}^{(N)} + \Delta\alpha_x\right) - E\left(G_{ain}^{(N)}\right)\right]^R\right). \end{aligned} \quad (34)$$

## 4 Proposed Approach

The proposed model aims at forming a sub-network out of LoRaWAN network servers which ultimately results in a LoRaWAN virtual cloud. Using the virtual cloud, the PSCs are prioritized without affecting the general services offered through the LoRaWAN setup even in the absence of a general cellular network.

From the LoRaWAN- assisted setup explained previously, the virtual clouds are focused on resource allocation while satisfying the conditions given in Eqns. (33) and (34). The proposed model solves the resource allocation problems over a self-enforcing agreement [22] by using the principles of game theory. These principles are defined similarly to job-migration, which takes a decision on whether to shift a service or not. The modeled game-theory-based job migration is resolved on the basis of decay rule dominance that helps for efficient allocation of resources.

---

**Algorithm 1.** Selection of available resources
1: Input: $\lambda$, $E$, $B$, $\alpha$, $\beta$, $T_{max}$
2: Output: $B'$
3: Initialize system model
4: i=1
5: While ($i \leq |B|$) do
6:　　Calculate $D_i^{(e)}, \lambda_i$
7:　　If $\left(\left(\frac{1}{\lambda_i}\right) \geq \min(t) \| \left(D_i^{(e)} = \min\right)\right)$
8:　　　　Mark network server as available
9:　　　　Store Id in $B'$
10:　　Else
11:　　　　Mark network server as unavailable
12:　　End if
13:　　i=i+1
14: End while

---

Let $D_i^{(e)}$ be the decay rate of each involved sub network server that forms a cloud setup with the core network server, such that $i \leq |B|$. Now, the game is very simple, any network server that has minimum decay rate wins and hosts the PSC application provided that constraints on memory and energy are held and the PSC load handled by LoRaWAN setup is less than or equal to its permissible strength. Once the decay rate is available for expected application, the proposed model uses Nash equilibrium [23] to attain a solution for the defined resource allocation (or job migration) problem.

Let $B'$ be the set of available network servers divided on the basis of decay rate such that their average availabilities are subject to their lifetime, which means a network server is available if $\frac{1}{\lambda_i}$ is maximum. The lifetime is accounted either on the basis of memory or energy from the formula $D_i^{(e)} = D_0 e^{-\lambda t}$ where $D_0$ can be understood with the help of an Algorithm 1. The maximum lifetime is accounted on the basis of the minimum time required by an incoming application let alone the energy or the memory requirements. The proposed model can be operated strictly by modifying the "IF" clause of the algorithm as "IF $\left(\left(\frac{1}{\lambda_i^{(\alpha)}} \geq \min(t)\right)$ and $\left(\left(\frac{1}{\lambda_i^{(e)}} \geq \min(t)\right)\right)$".

This helps to select the network servers that can obey the conditions of energy and memory requirements. Now, the problem can be modeled into Nash equilibrium [23] as shown in Figure 4. The proposed solution aims at selecting a state of the network where $E_R$ and $\alpha_R$ are maximum with $\alpha_R \geq \alpha_{TH}$ and $E_R \geq E_{TH}$. Here, $\alpha_{TH}$ and $E_{TH}$ are the thresholds for memory and energy, respectively, required for the continuous operations of the network. These are calculated from Eqns. (22) and (25) by inputs only for LoRaWAN traffic and LoRaWAN sever depletion rate when no PSC application is transferred to them.

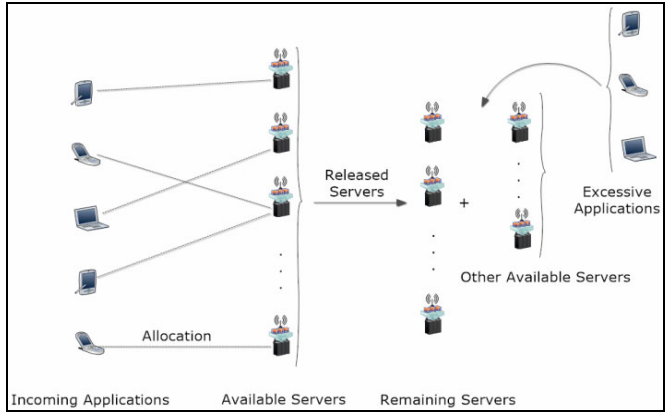

**Figure 4.** An illustration of the proposed solution for the defined resource allocation problem

Now, the applications are transferred to the available server in the First Come First Serve (FCFS) [42] order provided that the equilibrium conditions are satisfied, which include:

$$E_R \geq E_{TH}, \ \alpha_R \geq \alpha_{TH}, \forall B',$$
$$E_0^{(B')} e^{-K_B T_{max}} \geq \frac{1}{\delta_X} E_0^{(B)} e^{-K_B T_{max}},$$
$$\alpha_0^{(B')} e^{-K_B T_{max}} \geq \frac{1}{\delta_Y} \alpha_0^{(B)} e^{-K_B T_{max}}.$$

(35)

where $E_0^{(B')}$ and $\alpha_0^{(B')}$ are the total energy and memory of all the available network servers at the initial instance. $K_B \cdot$ and $K_B$ are the total applications on available servers and entire network prior to a disaster,[2] respectively. $E_0^{(B)}$ and $\alpha_0^{(B)}$ are the total energy and memory of the entire network prior to a disaster. $\delta_X$ and $\delta_Y$ are the instances for energy and memory resources, respectively, after which the required energy and memory conditions can be satisfied by the incoming applications at the LoRaWAN network servers. Thus, attainment of these equilibrium conditions duration job migrations helps to optimally allocate resources over the self-enforced agreement in LoRaWAN- assisted networks. All these procedures are provided as easy to follow steps in Algorithm 2.

---

**Algorithm 2.** Resource allocation for LoRaWAN- assisted network using Nash Equilibrium

1: Input: *N, S, B, $K_B$, $K_{B'}$, λ, E, α, β, $T_{max}$*
2: Output: *Remaining PSC applications = NILL*
3: Initialize system model and network model
4: While ((35)! =False) do
5:     Calculate energy and memory requirements of each application
6:     Calculate remaining energy and memory of available servers
7:     Allocate each application in FCFS order by following precedence in memory or energy
8:     If ((35) holds)
9:         Exit and wait for next pool of PSC
10:     Else
11:         Adjust thresholds and recheck conditions
12:         If (PSC handled = true)
13:             Mark iteration, exit and wait
14:         Else
15:             Allocate applications to servers with max. lifetime w.r.t. energy or memory
16:             Check conditions and exit if satisfied
17:         End if
18:     End if
19: End while

---

Algorithm 2 helps to iteratively allocate the resources as per availability and any continuous loop without outputs marks the deadlock or bottleneck. At the moment, only thresholds are used to exit from such a condition. However, in our later reports, emphasis will be given to recover the proposed solution from a possible deadlock using predictive mechanisms.

## 5 Performance Evaluations

The proposed approach is numerically analyzed for its performance by using MATLAB[TM]. The evaluations are performed by comparing the proposed approach with a baseline scenario, which is the state when all the entities in the network are able to support transmission without any failure. The network is operated at 50% and 100% failure rates of the existing infrastructure and the entire PSC traffic is regulated by the LoRaWAN gateways. The network is operated with a single core network server with a maximum of 100 APs, 10 LoRaWAN network servers that support 100 LoRaWAN gateways with 1000 users. The details of other metrics configured for evaluations are given in Table 1.

The initial evaluations are performed to calculate the number of links that are active irrespective of the induced failures in the network. The results show that the proposed approach provided 261 and 211 number

---

**Table 1.** Parameter configurations

| Symbol | Value | Description |
|---|---|---|
| $\lvert N \rvert$ | 1 | Number of core network servers |
| $\lvert B \rvert$ | 10 | Number of sub-network LoRaWAN servers |
| $\lvert L \rvert$ | 100 | Number of LoRaWAN gateways |
| $\lvert E \rvert$ | 500-1000 | Number of end users |
| $\lvert S \rvert$ | 100 | Number of APs |
| $X$ | 50% - 100% | Number of failed APs |
| $T, T_{max}$ | 3600, > 3600 s | Maximum running time |
| $t$ | 360 s | Time step divisions |
| $K$ | 10-100 per server | Number of applications |
| $\lambda$ | Distribution over $K$ | Arrival rate of applications |
| $\alpha$ | 1 – 10 MB | Memory consumed per application |
| $\alpha_0$ | 10,000 MB | Initial memory |
| $E_0$ | 0.0337 J | Initial energy available for each application on a server |
| $P_c t$ | 33.724 μJ | Energy consumed by each application in one second[3] |
| $L_T$ | 1211 | Total number of links |
| $n_1, n_2$ | 0.5 and 0.3 | Balancing constant for energy and memory |
| $K'$ | 10% w.r.t. $\lambda$ | Number of applications closing at each server |
| $B'$ | 1-5 | Number of left out servers |

of active links at 50% and 100% failure rate of APs compared with 311 connections of baseline at $\lambda$=100, as shown in Figure 5. The number of links helps to understand the load which will be shifted to LoRaWAN gateways in the absence of traditional APs, which otherwise are responsible for handling traffic for PSCs.

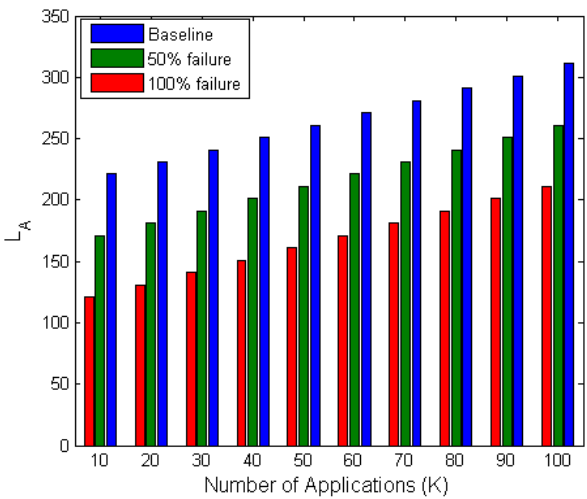

**Figure 5.** Number of active links vs. number of applications

The operational efficiency of the proposed setup lies in the probability of continuity, which helps to understand whether the network can sustain for the intended duration or not. A network with extremely lower probability is more likely to shut down causing tremendous loss to the traffic and resulting in an increasing number of unhandled users.

However, with the proposed self-enforcing agreement based model, the proposed approach is able to balance the load even in the absence of APs, which ultimately results in a high probability of connectivity for the entire duration of transmission even with the increasing number of applications at different failure rates. Results show that the proposed approach is able to provide an average probability of 0.93 for the baseline, whereas, for 50% and 100% failures of APs, this value decreases slightly to 0.92 only, as shown in Figure 6. This demonstrates the effectiveness of the proposed model in efficiently supporting the PSC applications.

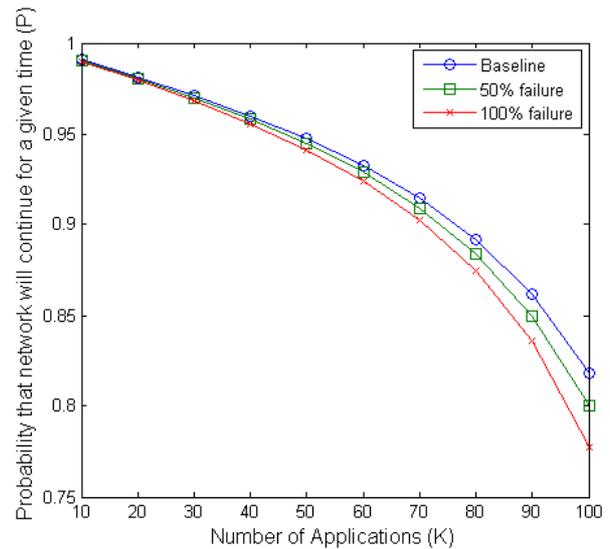

**Figure 6.** Probability of continuity for a given time vs. number of applications

Similar trends are observed for the probability to continue at the given configurations as shown in Figure 7. Results show that once the network is overloaded, the proposed approach is able to withstand the burden by providing sufficient resources to each application resulting in an average probability of 0.43 and 0.36 for 50% and 100% failing scenario, respectively, whereas for the baseline this value is 0.5, which show that at given configurations, the baseline is also under the threat of 50% failure. From such results, it can be noticed that average survivability of the proposed approach is high compared with the baseline conditions.

---

[3] https://www.cooking-hacks.com/documentation/tutorials/lorawan-for-arduino-raspberry-pi-waspmote-868-900-915-433-mhz [last accessed on February 2018]

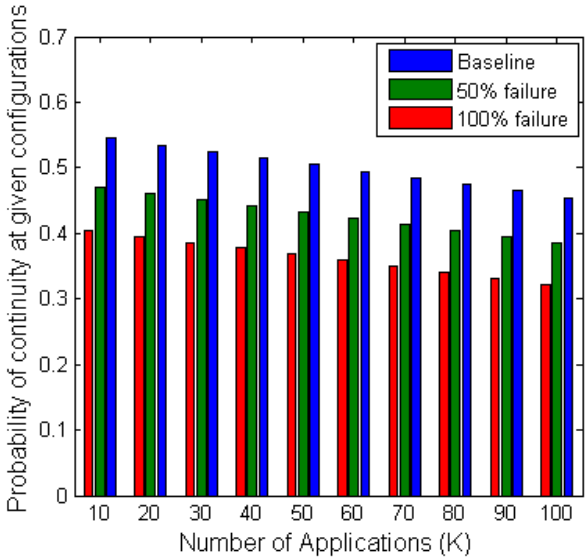

**Figure 7.** Probability of continuity at given configurations vs. number of applications

With more number of applications closing with time, the number of network resources is restored which helps to facilitate more applications. However, restoration is dependent on the number of links that are active and the recoveries of energy and memory units. Once maximum applications handled by the designated LoRaWAN servers are closed, the number of currently active applications decreases, which results in conservation of network resources leading to a higher value for network resource restoration as shown in Figure 8. Results show that the baseline scenario is able to provide a restoration value of 0.024 and that obtained for 50% and 100% failures of APs are 0.020 and 0.016, respectively. These values are competitive considering that the network can continue to operate even in the absence of its core service supporters. This helps to enhance the mean lifetime of each application by 0.37 seconds which otherwise would have been zero.

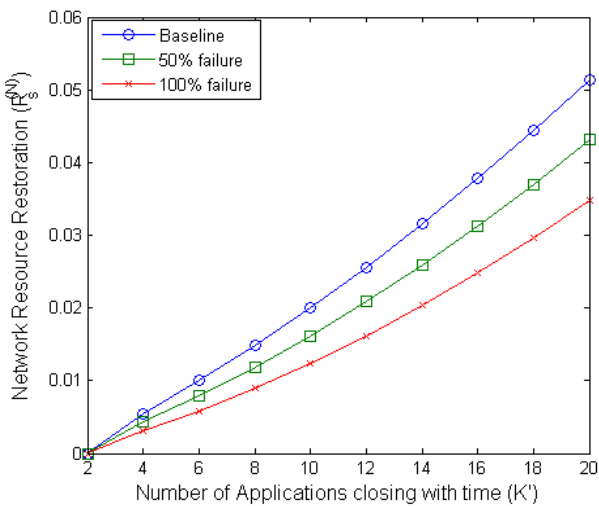

**Figure 8.** Network resource restoration vs. number of applications closing with time

In the proposed model, another important metric for driving the entire solution is the gain, which is observed for both the memory as well as the energy. With the same rate of operations, the trends for gains for both memory and energy are same with an equivalent slope as shown in Figure 9.[4] These results show that the overall gains decrease with an increase in the number of applications as more incoming applications consumes more memory and requires much energy for evaluations resulting in a decrease in the amount of memory and energy saved for each application. The results show that the proposed approach provides an average energy gain of 46.34%, 43.87%, and 41.17% for the baseline scenario and the scenarios with 50% and 100% failures, respectively.

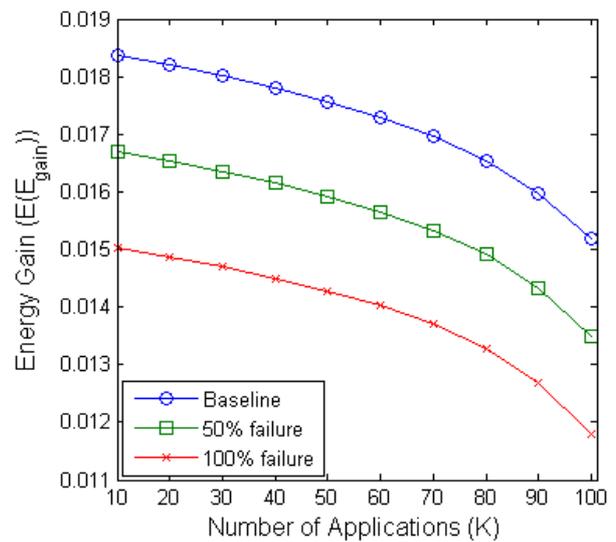

**Figure 9.** Energy gain vs. number of applications

In the proposed approach, the amount of available energy decides the excessive load that can be handled by the LoRaWAN servers, which is coming from the traditional setup of public safety network. It is conclusive that the amount of energy will increase with an increase in the number of incoming applications, and contrary to this, the amount of energy conserved will increase with an increase in the number of applications closing in a given instance. The results show that the amount of energy remains same for each application and thus, the actual conserved energy is also same for all the three scenarios even at increasing number of closing applications as shown in Figure 10. The proposed approach is able to conserve an average energy 0f 0.011 J for each closing application irrespective of the server hosting it.

---

[4] Because of similar slope and trend, graph only for energy gain is presented.

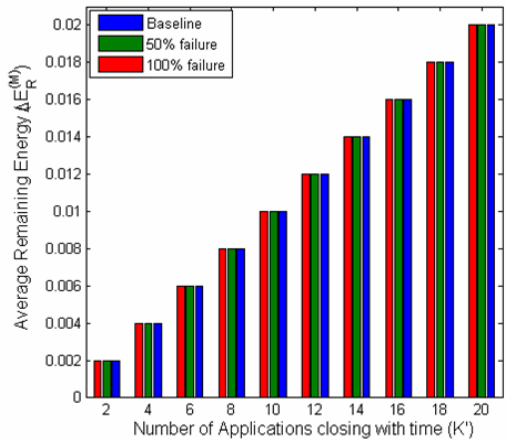

**Figure 10.** Average remaining energy vs. number of applications closing with time

All the factors proposed in this paper are studied w.r.t. self-enforcing agreement between the metrics, such as memory and energy for a given set of components. The enforcing conditions in (10) and (20) are used for deciding the allocation of resources across the available servers and also for the resource restorations. It is evident from the previous results on energy and memory conservation that the proposed approach is able to handle excessive load provided that there are sufficient resources available at the LoRaWAN network servers. Even if the resources are limited compared with the incoming applications, the proposed model allocates them effectively resulting in zero-halt time for all the PSC applications. Results, as shown in Figure 11, show that the proposed approach in the baseline mode shows better self-enforcing criteria as this is very much likely to occur and decreases to a slightly lower value for 50% and 100% failures. Results show that the proposed approach shows a variation of only 14.40% and 27.28% for 50% and 100% failures in comparison with the baseline scenario. From these results, it is evident that the proposed approach can efficiently support resource allocation for PSC application by using LoRaWAN setup in the absence of traditional APs.

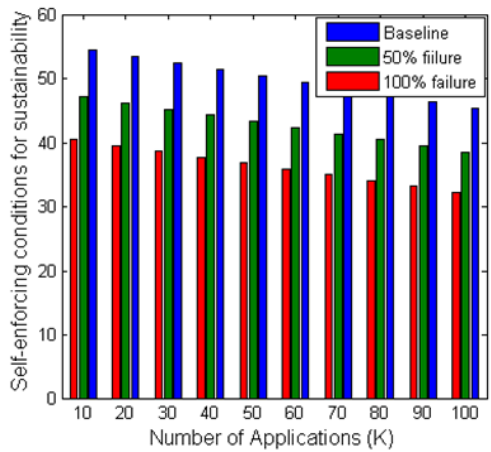

**Figure 11.** Self-enforcing values for sustainability vs. number of applications

# 6 Discussions and Comparisons

The proposed approach uses self-enforcing agreement [22] and Nash equilibrium algorithm [23] for satisfying the conditions imposed by the policies over memory and energy requirements of the incoming applications. The proposed approach successfully formulates the optimization criteria which are taken care of while allocating applications to the LoRaWAN cloud formed in the absence of the traditional setup, which is responsible for supporting PSC applications. The LoRaWAN cloud helps to disseminate the traffic across multiple LoRaWAN sub-network servers that support PSC traffic without affecting its regular operations as well as the lifetime of the network. There are a large number of approaches that have focused on the resource allocation issues over different types of setups but considering LoRaWAN architecture, there is a scarcity of solutions that have utilized its features while forming energy-efficient as well as highly sustainable and survivable network model. This paper extends the state-of-the-art solutions available for enhancing the energy efficient resource allocations by utilizing the memory and energy agreements between the LoRaWAN servers and the operational APs for PSCs.

Despite the fact that there is a limited number of solutions which have utilized LoRaWAN-based cloud setups for PSCs, this papers highlight come of the key contributions for energy-efficient resource allocation and compare them with the proposed solution as shown in Table 2. From the comparisons, it is convincingly evident that the proposed approach is novel in terms of the LoRaWAN-cloud formation while supporting the PSCs in absence of traditional APs. Also, it supports the combination of energy-efficient resource allocation as done by most of the state-of-the-art solutions, but with unique network formations and light-weight mechanisms for decisions on the shifting of applications with sufficiently high survivability and sustainability. In future, the proposed approach can be facilitated with privacy-preserving resource provisioning solutions for the virtual cloud formed by the LoRaWAN servers, and user-attribute-based allocation schemes can be adopted for securing PSCs applications over these servers [47-50]. Resource allocation in PSCs can be operated via offloading paradigms which help to attain a low-overhead scenario during the transmissions of critical information [51-53]. Further, these mechanisms can be improved by using approaches like network slicing, digitalized operations and centralized control through Software-Defined Networking and Network Function Virtualizations [54-55].

**Table 2.** Comparison with the state-of-the-art solutions for energy-efficient resource allocation

| Approach | Ideology | Application | Resource Allocation | LoRaWAN-based solution | Energy-Efficient | Survivability and Sustainability |
|---|---|---|---|---|---|---|
| Parking Bay Management [43] | LoRaWAN for parking system | - | No | Yes | Yes | No |
| UAVs for PSC [35] | Capacity enhancement and coverage | PSC | - | No | - | No |
| Energy Efficient Resource Allocation [44] | Mobile edge computation offloading | Mobile Edge Networks | Yes | No | Yes | - |
| Energy Efficient Resource Allocation [45] | Resource allocation as stochastic network optimization | Buffer-aided wireless relay networks | Yes | No | Yes | - |
| Energy Efficient Resource Allocation [46] | Resource allocation via successive convex approximation | Augmented reality applications | Yes | No | Yes | - |
| Proposed Approach | Self-enforcing agreement on LoRaWAN | PSC | Yes | Yes | Yes | Yes |

## 7 Conclusions

This paper proposed a novel network model via a combination of LoRaWAN and traditional public safety networks and used self-enforcing agreement for allocating the resources efficiently amongst the available servers. These agreements were formulated on the basis of energy and memory constraints and the solution for agreements was attained using Nash equilibrium. The proposed approach was highly supportive in handling an excessive number of applications in case the traditional Access Points (APs) fail to accommodate the incoming Public Safety Communication (PSC) requests.

The proposed approach was numerically evaluated and compared with a baseline scenario with no failure of APs. The results show that the average probability of connectivity for the proposed approach is 0.92 in case of dependency on LoRaWAN servers compared with 0.93 of the baseline scenario. The mean lifetime of each application is enhanced by 0.37 s which is zero in case of a network failure. Further, the proposed approach provides an average gain of 46.34%, 43.87%, and 41.17% for the baseline scenario and the scenarios with 50% and 100% failures, respectively, for both memory and energy constraints of the network. Finally, in self-enforcing agreements, the proposed approach shows a variation of only 14.40% and 27.28% for scenarios with 50% and 100% failures of APs, respectively, in comparison with the baseline network with no failure of nodes.

## Acknowledgment

This work was supported by Institute for Information & communications Technology Promotion (IITP) grant funded by the Korea government(MSIP) (2015-0-00508, Development of Operating System Security Core Technology for the Smart Lightweight IoT Devices) as well as the Soonchunhyang University Research Fund.

## Biographies

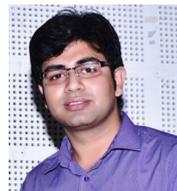
**Vishal Sharma** received the Ph.D. and B. Tech. degrees in computer science and engineering from Thapar University (2016) and Punjab Technical University (2012), respectively. He worked at Thapar University as a Lecturer from April 16-October 16. Dr. Sharma was a postdoctoral researcher at Soongsil University and Soonchunhyang University, South Korea from November 16-September 17. Now, he is a Research Assistant Professor at Department of Information Security Engineering, Soonchunhyang University, Republic of Korea.

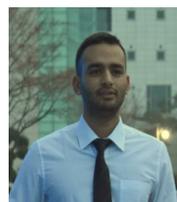
**Gaurav Choudhary** received the B. Tech. degree in Computer Science and Engineering from Rajasthan Technical University in 2014 and the Master Degree in cyber security from Sardar Patel University of Police in 2017. He is currently pursuing Ph.D. degree in the Department of Information Security Engineering, Soonchunhyang University, Asan, South Korea.

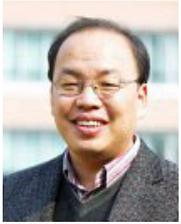 **Ilsun You** received the M.S. and Ph.D. degrees in computer science from Dankook University, Seoul, Korea, in 1997 and 2002, respectively. He received the second Ph.D. degree from Kyushu University, Japan, in 2012. From 1997-2004, he was at the THIN multimedia, Internet Security, and Hanjo Engineering as a research engineer. Now, he is an associate professor in Information Security Engineering Department, Soonchunhyang University. He is a Fellow of the IET.

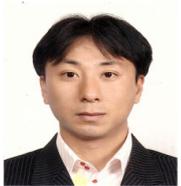 **Jae Deok Lim** received his M.S. degree in Electronic Engineering from Kyungbook National University in 2001 and his Ph.D. degree in Computer Engineering from Chungnam National University in 2013. He is currently a senior researcher at Information Security Research Division in ETRI (Electronics and Telecommunications Research Institute). His research interests include IoT security, access control and secure operating system.

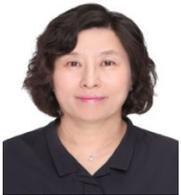 **Jeong Nyeo Kim** received her M.S. degree and Ph.D. in Computer Engineering from Chungnam National University, Rep. of Korea, in 2000 and 2004, respectively. She studied computer science from the University of California, Irvine, USA in 2005. Since 1988, she has been a principal member of engineering staff at the Electronics and Telecommunications Research Institute (ETRI).